\documentclass[a4paper,fleqn,usenatbib]{mnras}

\usepackage[T1]{fontenc}
\usepackage{ae,aecompl}
\usepackage{graphicx}

\usepackage{graphicx}	
\usepackage{amsmath}	
\usepackage{amssymb}	

\title[Plasmoid instability in a confined flare]{\textit{The plasmoid instability in a confined solar flare}}

\author[D. MacTaggart and L. Fletcher]{
David MacTaggart,$^{1,2}$\thanks{E-mail: david.mactaggart@glasgow.ac.uk}
and Lyndsay Fletcher$^{3,4}$
\\
$^{1}$School of Mathematics and Statistics, University Place, University of Glasgow, Glasgow, G12 8QQ, UK.\\
$^{2}$Department of Mathematics, 38123 Povo, University of Trento,  Italy.\\
$^{3}$School of Physics and Astronomy, Kelvin Building, University of Glasgow, Glasgow, G12 8QQ, UK.\\
$^{4}${Rosseland Centre for Solar Physics, University of Oslo, P.O.Box 1029 Blindern, NO-0315 Oslo, Norway.}
}

\date{Accepted XXX. Received YYY; in original form ZZZ}

\pubyear{2019}

\begin{document}
\label{firstpage}
\pagerange{\pageref{firstpage}--\pageref{lastpage}}
\maketitle

\begin{abstract}
Eruptive flares (EFs) are associated with erupting filaments and, in some models, filament eruption drives flare reconnection. Recently, however, observations of a confined flare (CF) have revealed all the hallmarks of an EF  (impulsive phase, flare ribbons, etc.) without the filament eruption itself. Therefore, if the filament is not primarily responsible for impulsive flare reconnection, what is? In this Letter, we argue, based on mimimal requirements, that the \emph{plasmoid instability} is a strong candidate for explaining the impulsive phase in the observed CF. We present magnetohydrodynamic simulation results {of the nonlinear development of the plasmoid instability, in a model active region magnetic field geometry,} to strengthen our claim. We also discuss how the ideas described in this Letter can be generalised to other situations, including EFs. 
\end{abstract}




\section{Introduction}
In the leading models of solar eruptive flares (EFs), such as the \emph{standard flare model for a two-ribbon flare} \citep{1964NASSP..50..451C, 1966Natur.211..695S,1974SoPh...34..323H,1976SoPh...50...85K}, {its extension into 3D} \citep[e.g.][]{aulanier12,aulanier13,janvier13} and the \emph{breakout model} \citep[e.g.][]{antiochos99,karpen12}, flare reconnection is associated with filament eruption. Although different models {for EFs emphasise different mechanisms allowing the onset of} eruptions, they all contain filaments {or magnetic filament channels} whose magnetic geometry allows flare reconnection to occur in a vertical current sheet. {Flares which do not exhibit evidence for eruptions are known as \emph{confined flares} (CFs) \citep{1977ApJ...216..108P}. Whereas in an EF the filament/filament channel eruption is a key factor, in a CF this is not the case}. Recently, \cite{simoes15a,simoes15b} studied a bipolar active region that contained a filament. This region produced an impulsive (C-class) flare in the corona and the formation of two flare ribbons. The filament of the region, however, did not erupt.  

In order to understand the large-scale behaviour of the CF, we need to move beyond the models of EFs. {Theories for CFs have not attracted so much attention, though prominent examples include those based on the interaction of current-carrying loops \citep[e.g.][]{1997ApJ...486..521M} and the interaction of emerging flux with pre-existing coronal fields \citep[e.g.][]{heyvaerts77}.}
In this Letter, we present a {new} theoretical description of how the impulsive phase of the CF forms. We note that CFs need not have a unique form or magnetic topology, so that what we will describe in this Letter may not cover all cases of CFs. However, the region described in \cite{simoes15a,simoes15b} is a bipolar region which is the basic building block of solar active regions \citep{schrijver00} and, therefore, a very general structure in the solar atmosphere.  

In the following section, we describe the general features of our theory. We then present magnetohydrodynamic (MHD) simulation results to strengthen our claim. We conclude the Letter with a summary and a discussion of how the ideas discussed can be applied to a more general setting. 

\section{General theoretical description}
In this section, we describe the main qualitative features of the theory. Since we are interested in structures that exist on active region length scales, we will model the plasma using resistive MHD. The general properties of the theory, however, are not strictly dependent on MHD and could be extended to non-fluid theories of plasmas. 

In the region studied by \cite{simoes15a,simoes15b} there is a filament which does not erupt {during the observed flare}. Therefore, in order to understand how an impulsive phase could develop, we ignore the direct influence of the filament. One constraint from the observations is that the impulsive phase has a coronal source, probably at the top of the active region's magnetic field. 

Leaving the dynamics of the filament aside, we are left with two distinct magnetic domains - that of the bipolar active region and that of the corona surrounding the active region. These two domains are not in equilibrium, otherwise there would be no subsequent flaring. Let us assume, however, that any {evolution} behaves in a quasi-static fashion, i.e. there are no strong motions to drive impulsive flare reconnection. 

A \emph{current sheet} exists at the boundary between the two magnetic domains. Here, a current sheet refers to a thin, but finite, layer of {locally enhanced current density.}

The idea of  flares resulting from the interaction of active region and coronal fields is not a new one in solar physics \citep[e.g.][]{heyvaerts77}. However,  our theory represents a {substantial} `update' of this established idea. Consider the slow movement of the active region domain. It is possible that some locations of the current sheet can be compressed more than others, especially as the relative orientation of the magnetic field domains near the boundary will be different in different locations on the boundary (in 3D). Assuming that part of the current sheet continues to be compressed, it can do so until a critical aspect ratio (sheet thickness over sheet length) is reached. Beyond this point, the current sheet becomes unstable to the \emph{plasmoid instability} \citep[e.g.][]{loureiro07,pucci14,uzdensky16}. Why is this instability suitable for describing the impulsive phase of a flare? Firstly, the linear phase of the plasmoid instability is very fast. For a current sheet with the Sweet-Parker scaling, the {linear} growth rate of the instability is $O(S^{1/4})$ where $S$ is the \emph{Lundquist} number based on the macroscopic length of the current sheet\footnote{Even if the Sweet-Parker scaling cannot be reached, the growth rate is still a positive power of $S$ and thus fast \citep{pucci14}.}. A typical coronal estimate is $S\sim 10^{13}$ \citep[e.g.][]{comisso17}. 

Secondly, the plasmoid instability leads to the formation of many {highly dynamic} plasmoids {in its nonlinear phase} \citep[e.g.][]{lapenta08,bhattacharjee09,samtaney09,tenerani15,huang17} and, therefore, many new locations of magnetic reconnection. This fact is important for flares as each location of reconnection is also a location of enhanced parallel (to the magnetic field) electric field \citep[e.g.][]{schindler88}. Therefore, {assuming that plasmoid formation cascades so that sufficient numbers form at sufficiently small scales, as found in the high resolution 2D simulations cited above,} there are many regions in the current sheet where particles can be accelerated multiple times, reaching higher energies \citep[e.g.][]{2006Natur.443..553D,turkmani06,zhou18}.

The above two properties make the plasmoid instability a strong candidate for impulsive flares in CFs.  The only requirement is a thinning current sheet driven by slow motions due to a lack of equilibrium. We will now show an example of this in a 3D MHD numerical experiment.  

\section{Simulation}
So far, we have given a qualitative description of the impulsive phase of {a  bipolar CF}. 
Our argument has been based on some basic plasma physics and is not difficult to generalise to more complex situations (we will return to this later).  In order to add weight to the possibility of our theory being realised, we now present an analysis of a simulation that was originally described in \cite{dmac15}. Although magnetic topology is discussed in that work, the main focus is the formation of surges. We, therefore, go back to this simulation and analyse the data in light of our description of the CF.


{The compressible and resistive magnetohydrodynamic equations are solved using a Lagrangian remap scheme \citep{arber01}. In dimensionless form, the equations are 
\begin{equation}\label{mass_con}
{\dot{\rho}} = -\rho\nabla\cdot\mathbf{u},
\end{equation}
\begin{equation}\label{mom_con}
{\dot{\mathbf{u}}} = -\frac{1}{\rho}\nabla p + \frac{1}{\rho}(\nabla\times\mathbf{B})\times\mathbf{B}+\mathbf{g}+\frac{1}{\rho}\nabla\cdot\mathbf{T}_V,
\end{equation}
\begin{equation}\label{induction}
{\dot{\mathbf{B}}} = (\mathbf{B}\cdot\nabla)\mathbf{u} - {(\nabla\cdot\mathbf{u})\mathbf{B}} +\eta\nabla^2\mathbf{B},
\end{equation}
\begin{equation}\label{energy_con}
{\dot\varepsilon} = -\frac{p}{\rho}\nabla\cdot\mathbf{u} + \frac{1}{\rho}\eta j^2+ \frac{1}{\rho}\mathbf{T}_V:{\nabla\mathbf{u}},
\end{equation}
\begin{equation}\label{divB}
\nabla\cdot\mathbf{B} = 0,
\end{equation}
with specific energy density
\begin{equation}\label{energy_den}
\varepsilon = \frac{p}{(\gamma-1)\rho}.
\end{equation}
The over-dot represents the material derivative and the double-dot represents the double contraction of a second order Cartesian tensor. The basic variables are the density $\rho$, the pressure $p$, the magnetic field $\mathbf{B}$ and the velocity $\mathbf{u}$. $j$ is the magnitude of current density, $\mathbf{g}$ is {the gravitational acceleration} and $\gamma (=5/3)$ is the ratio of specific heats. $\eta$ is the resistivity and its value is taken to be, $\eta=10^{-3}$. This value can also be expressed  as the global Lundquist number based on the non-dimensional length scale, i.e. $S=1000$. 

The viscosity tensor is given by
\begin{equation}\label{visc_ten}
\mathbf{T}_V = \mu\left(\nabla\mathbf{u}+\nabla\mathbf{u}^{\rm T}-\frac{2}{3}\mathbf{I}\nabla\cdot\mathbf{u}\right),
\end{equation}
where $\mu=10^{-5}$ and $\mathbf{I}$ is the identity tensor. The non-dimensionalisation and setup of the initial condition is identical to that in Section 2 of \cite{dmac15} and we refer the reader to that work for further details. }
 
\begin{figure}
\includegraphics[width=\columnwidth]{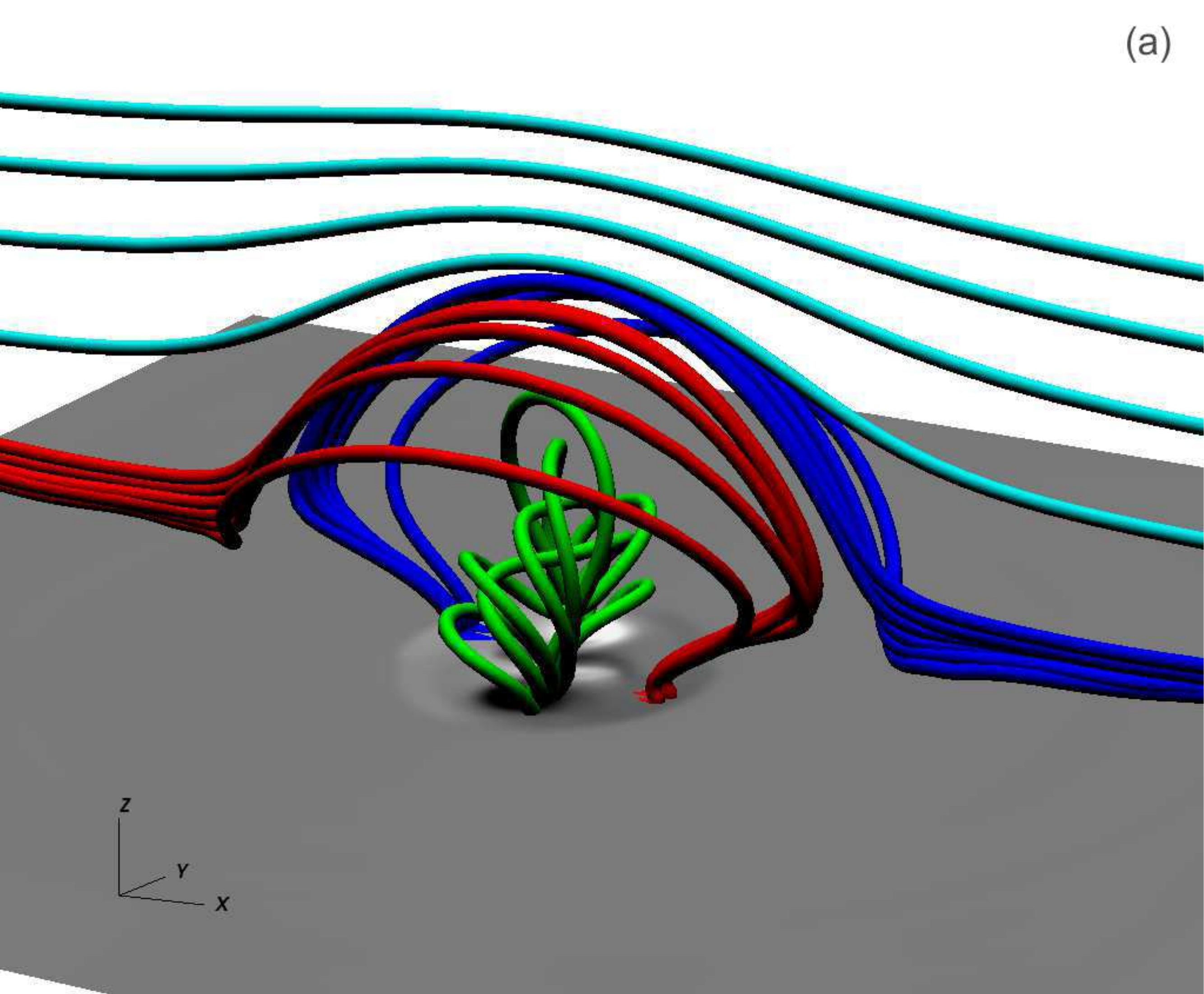}

\includegraphics[width=\columnwidth]{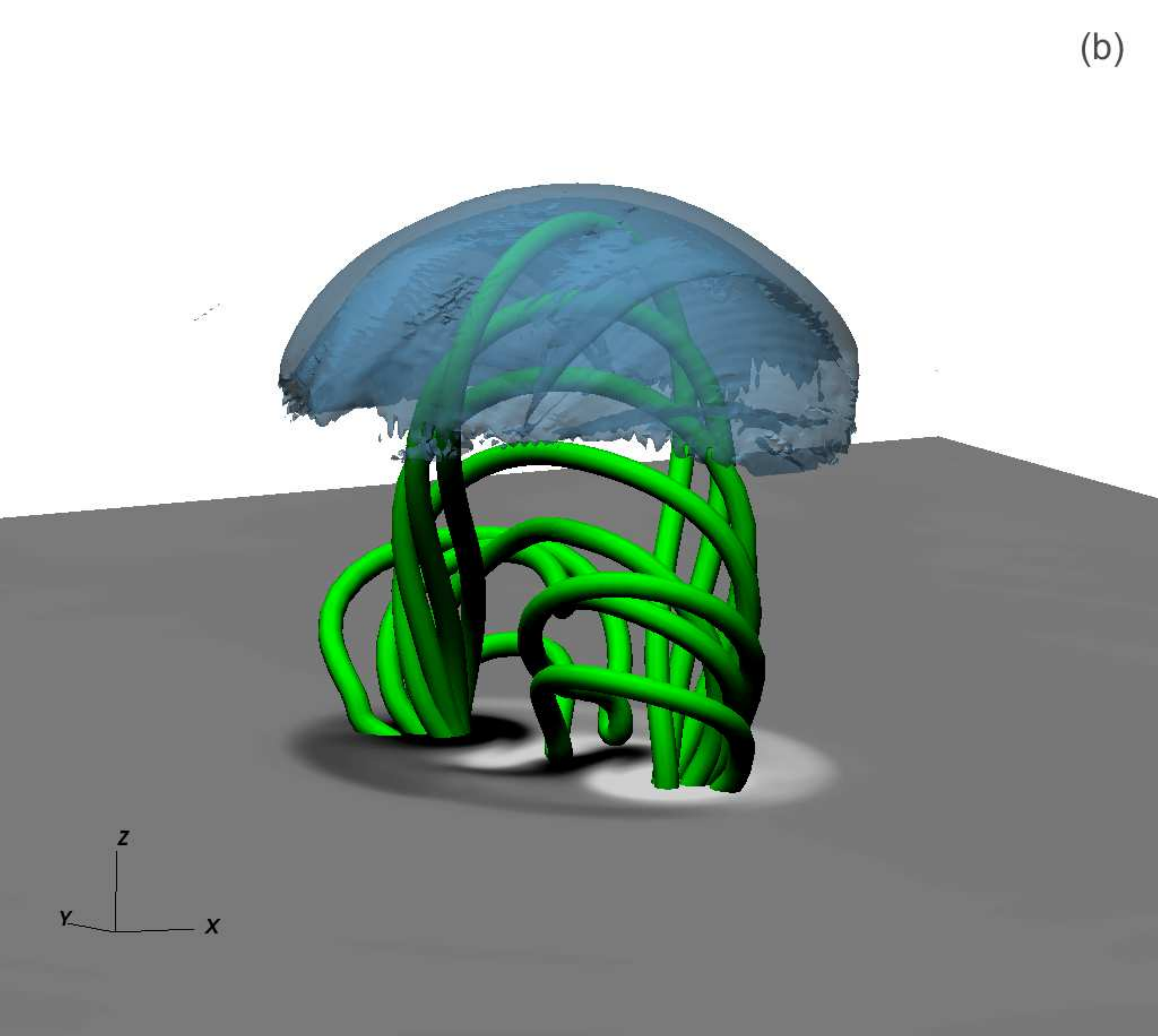}
\caption{The active region field and coronal field in the quasi-steady phase. {(a) shows the field line connectivity with reconnected field lines. (b) shows the isolated current sheet where the flare occurs.} Further details are given in the main body of the text.}
\label{steady}
\end{figure}

We ignore the initial emergence of the field into the atmosphere and skip to a later time when the overall kinetic energy of the simulation is decaying and a bipolar region with a filament has formed, surrounded by a coronal field. A visualisation of this scenario is displayed in Figure \ref{steady}.

The connectivity of field lines is labelled by colour {in Figure \ref{steady}(a)}. Green field lines connect to both active region footpoints. Cyan field lines connect to two  sides of the computational domain (the coronal field). The red and blue field lines are reconnected field lines and connect from one side of the computational domain (the corona) to one of the footpoints (at the base of the computational domain below the greyscale plane indicating the photosphere). A {simulated} magnetogram of the vertical component of the magnetic field is shown on the photospheric plane. {Figure \ref{steady}(b) displays the current sheet at the top of the active region (an isosurface of $j_x=0.001$).}

The magnetic field in Figure \ref{steady} corresponds to a time after emergence and before the impulsive flare. The shape of the active region `bubble' is clearly visible from the geometry of the reconnected field lines and the sheared filament is displayed (green field lines) in Figure \ref{steady}(a). There are two important points to consider about this magnetic structure. The first is that the reconnected field lines \emph{stay on the boundary of the active region}. Even at the photosphere, the reconnected field lines do not penetrate \emph{inside} the active region to connect to the main footpoints (sunspots). We will discuss how the behaviour of this quasi-static reconnection can change later.

The second point is that the reconnection occurs in a quasi-symmetric fashion, with the axis of symmetry lying approximately in the $x$-direction. This is indicative of a topological structure that appears in many different contexts. Dividing the four magnetic domains (indicated by the four colours), there is a magnetic separator \citep[e.g.][]{priest09}. This structure, and the ensuing reconnection, appears time and again in different studies. For example, in models of EFs \citep{dmac14}, in studies of reconnection at null points \citep{wyper14a}, in fly-by experiments \citep{parnell04,haynes07}, in magnetospheric studies \citep{dorelli07,haynes07b} and in an early model of flaring in \cite{sweet58} \citep[see also][and references therein for a detailed topological description of this last work]{longcope05}.  Interestingly,  2.5D \emph{resistive} MHD equilibria \citep{watson02,tassi03} in cylindrical coordinates have been found with a similar structure, including the presence of a separator. The existence of such solutions, combined with the frequency with which this magnetic structure occurs in different situations, suggests that the model representation (in Figure \ref{steady}) of this pre-flare active region described in \cite{simoes15a,simoes15b} is a general magnetic structure.

\begin{figure}
\includegraphics[width=\columnwidth]{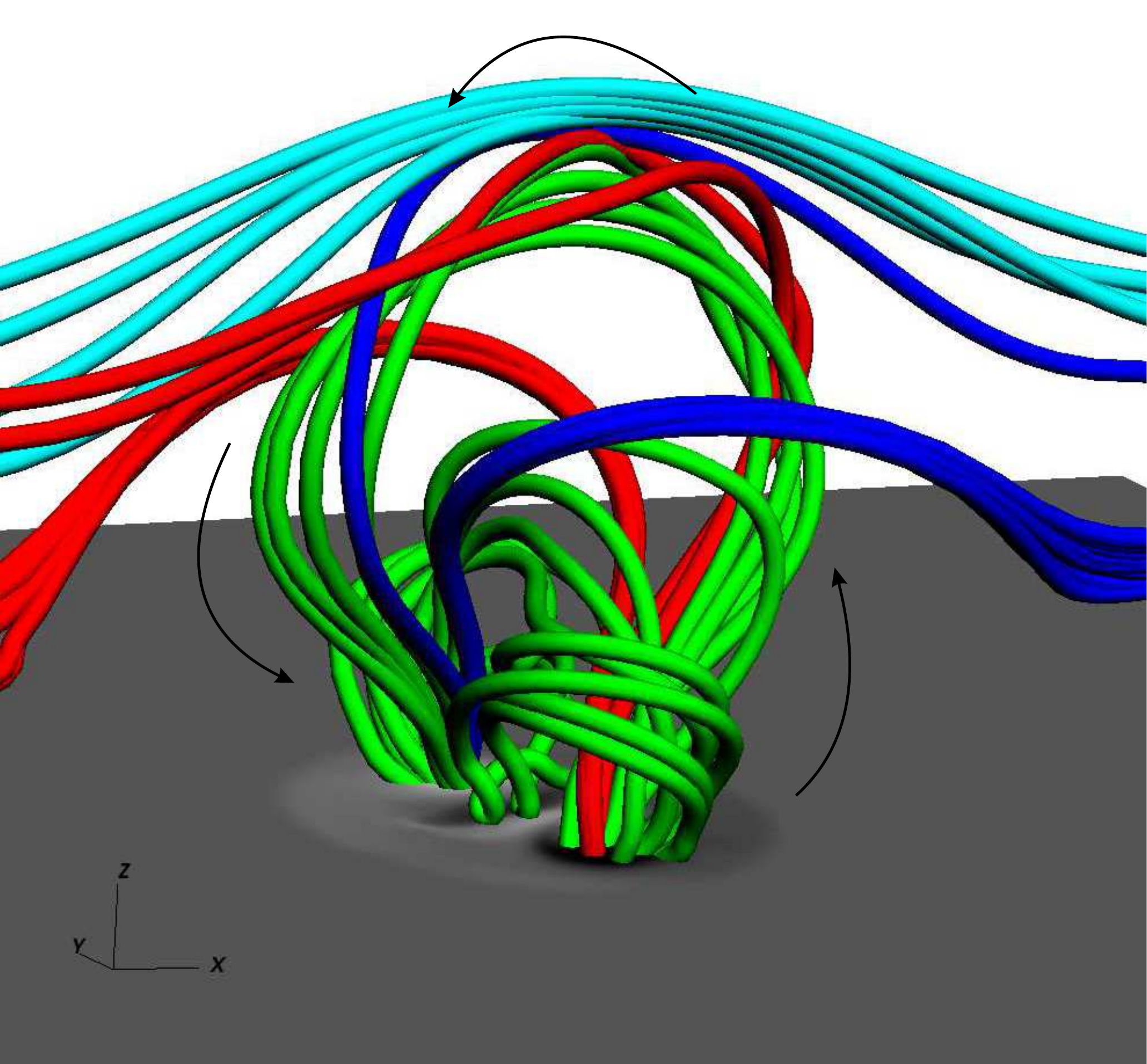}
\caption{The active region and coronal fields when tearing sets in at the boundary. {Arrows indicate directions for the description in the main text.}}
\label{tear}
\end{figure}

At a later time, the current sheet on the boundary (see Figure \ref{steady}(b)) eventually succumbs to tearing and the plasmoid instability. Figure \ref{tear} shows the magnetic topology after the onset of tearing.
The plasmoid instability is normally studied in 2D, where the change in topology is clear. In order to understand the connectivity displayed in Figure \ref{tear} begin from one footpoint, travel to the top of the active region while staying on the boundary and then travel to the other footpoint. {While doing so}, make note of the colour of the reconnected field lines. For example, let us start at the footpoint with negative magnetic field (shown in black {in the lower right-hand source} on the magnetogram). If we move upwards along the boundary, {following the arrows}, we first encounter blue reconnected field lines, then red, then blue and, finally, red again before we reach the other footpoint. Each red or blue region represents  {a plasmoid} - a topologically distinct region, created by the tearing of the boundary current sheet. This is the 3D realisation of the magnetic islands found in 2D studies of current sheet tearing. Of course, only a limited selection of field lines is displayed in Figure \ref{tear} for clarity.  

\begin{figure}
\includegraphics[width=\columnwidth]{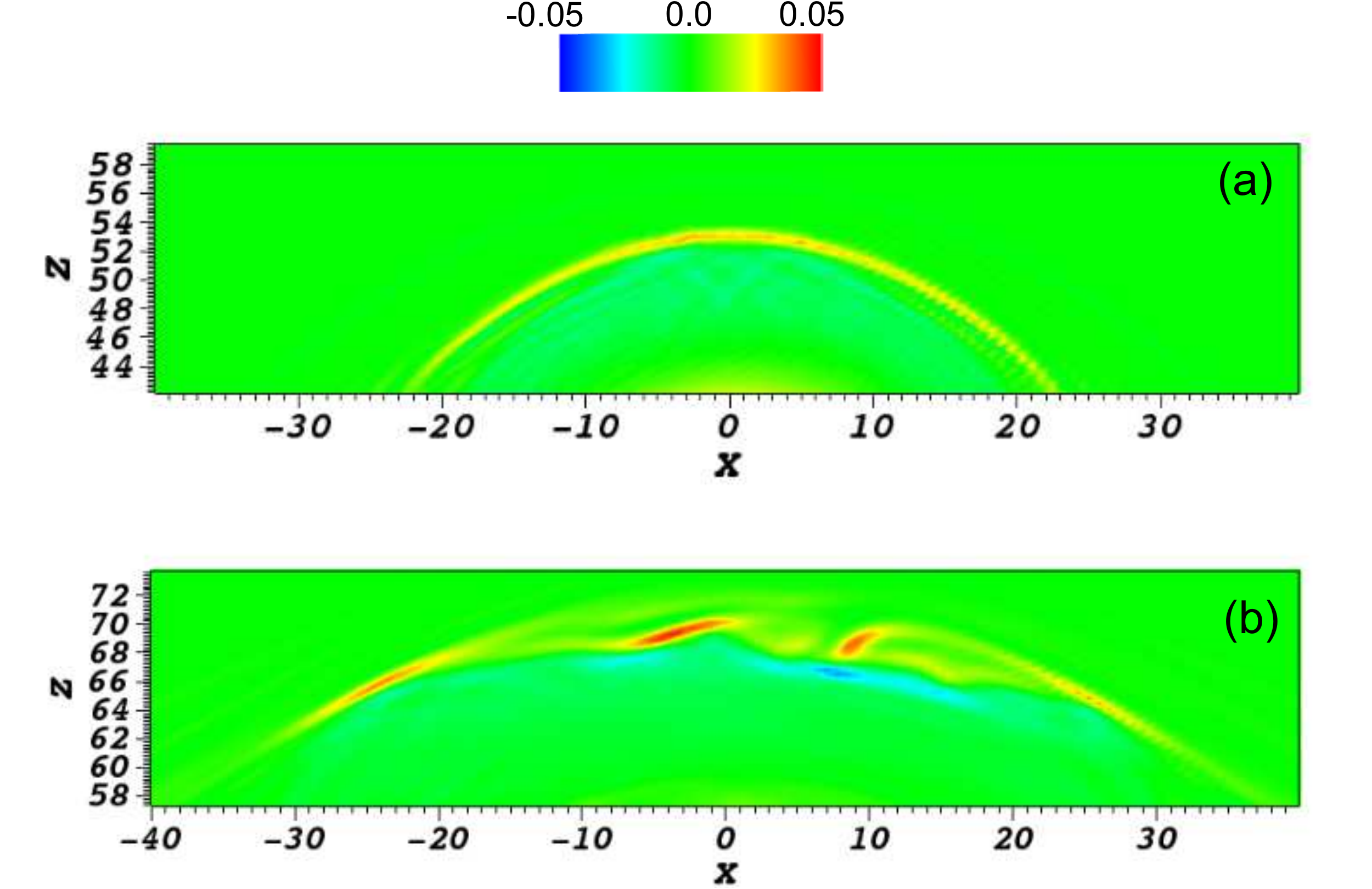}
\caption{{Slices in the $y=0$ plane of $j_y$. (a) corresponds to the state of the active region in Figure \ref{steady} and (b) to that in Figure \ref{tear}.} }
\label{slices}
\end{figure} 
{Figure \ref{slices} displays slices of $j_y$ in the $y=0$ plane for the snapshots shown in Figures \ref{steady} and \ref{tear}. Figure \ref{slices}(a) shows a thin but coherent current sheet with the peak current at the top of the active region, corresponding to where the quasi steady-state reconnection, shown in Figure \ref{steady}(a), occurs. Figure \ref{slices}(b) reveals the fragmentation of the current sheet from a coherent structure to three separated and intensified regions of current corresponding to the transitions between the plasmoids shown in Figure \ref{tear}. Notice that the current sheet in (b) is higher than that in (a). This increase in height is due to reconnection weakening the overlying tension of the coronal field, allowing the active region to move higher in the atmosphere.}  

\begin{figure}
\includegraphics[width=\columnwidth]{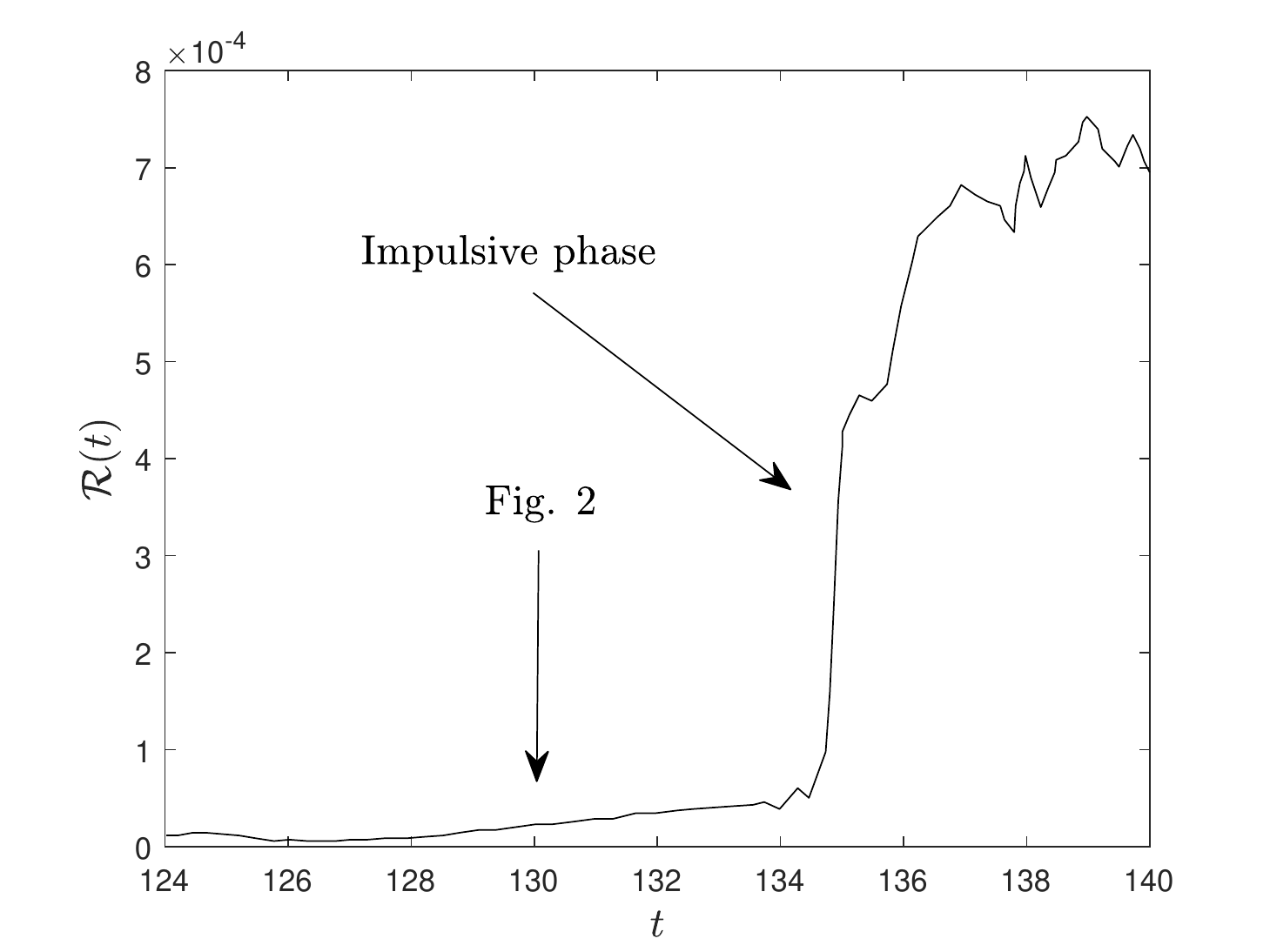}
\caption{The reconnection rate showing the impulsive phase. {These data were originally displayed in \citet{dmac15}}.}
\label{rate}
\end{figure}

The tearing of the boundary current sheet exhibits an impulsive phase, {as in the nonlinear phase of the plasmoid instability}, with the rapid creation of many plasmoids (the 3D definition of plasmoids described above). An alternative way to view the formation of many plasmoids is the bifurcation of the separator \citep{parnell10}. We define the 3D reconnection rate $\mathcal{R}(t)$ in the boundary current sheet as
\[
\mathcal{R}(t) =\max_{\mathcal{S}(t)}\left|\int_{\Gamma(t)}{E}_{\|}\,{\rm d}l\right|,
\]
where $E_{\|}$ is the component of the electric field parallel to the magnetic field, $\Gamma(t)$ is the path of the separator in the current sheet at time $t$ and $\mathcal{S}(t)$ is the set of all separators in the current sheet at time $t$. This definition of the reconnection rate is not unique but is suitable for revealing the impulsive phase.  Figure \ref{rate} shows how $\mathcal{R}(t)$ evolves from the state displayed in Figure \ref{steady} to that in Figure \ref{tear} and beyond. {The separators used in the calculation of the reconnection rate were found using the method of \cite{haynes07}.}

The qualitative details are of more interest than the quantitative details here, with the latter being discussed in \cite{dmac15}. For reference, however, Figure \ref{steady} is at $t=120$ ({multiply} by 25 to convert to seconds). This state remains until $t=129$ and Figure \ref{tear} is at $t=130$. After the tearing begins, there is a slow linear rise in $\mathcal{R}(t)$ until a fast transition followed by a higher saturated value. When tearing begins, the number of separators changes from 1 to 3, i.e $|\mathcal{S}|=3$. At the rapid transition, $|\mathcal{S}|=33$. In the saturated phase after the transition, $|\mathcal{S}|=1$, i.e. the system returns to having one separator.

The behaviour described above matches, qualitatively, the behaviour of the {nonlinear phase of the} plasmoid instability {\citep[e.g.][]{huang17}}. The system moves from a quasi-steady-state to the rapid formation of plasmoids boosting the reconnection rate and then back to a less dynamic state. 

The topology of the post-tearing field differs from that displayed in Figure \ref{steady}(a). Notice in Figure \ref{tear} that the reconnected field lines (blue and red) now connect \emph{inside} the active region (and inside the main footpoints) rather than skirting the edge of the domain (as in Figure \ref{steady}(a)). This new connection is possible due 3D reconnection, as has been exploited in models of EFs \citep[e.g.][]{janvier14}.

\section{Summary and Discussion}
In this Letter we have argued that the plasmoid instability exhibits properties consistent with the impulsive phase of confined flares in bipolar active regions.
The bipolar active region observed by \cite{simoes15a,simoes15b} produced a flare with an impulsive phase and other characteristic activity, such as ribbons. However, the filament in the region did not erupt. Without some kind of instability driving the filament upwards, as in models of eruptive flares, another explanation of the impulsive phase is required. Assuming no strong flows associated with the filament, we propose that since the two main flux domains (the active region and the surrounding corona) are not in equilibrium, compression of the current sheet between them (due to slow motion)  can eventually result in the plasmoid instability. Not only is the rapid nature of this instability suitable for describing the impulsive phase but it also occurs in the corona far from the photosphere, just as in the observations. 

{As evidence to support the possibility of the above theory, we have presented MHD simulation results of the nonlinear phase of the plasmoid instability in a realistic magnetic geometry for a solar flare. In particular, the simulation exhibits the following causal sequence of events:
\begin{itemize}
\item[1.] A thinning current sheet leading to reconnection.

\item[2.] The rapid formation of many plasmoids in the current sheet.

\item[3.] A highly boosted reconnection rate.

\end{itemize}}


The relationship between the plasmoid instability and solar flares has been suggested previously \citep[e.g.][]{uzdensky16,comisso17,janvier17}. However, to our knowledge, there has not previously been a 3D model or simulation of a flaring active region (erupting or non-erupting) which explicitly recognises the importance of the plasmoid instability.  This work extends the applicability of the plasmoid instability from 2D to 3D.

Another important aspect of this work is that the basic magnetic topology involved and its bifurcation (via the plasmoid instability) appear to be common phenomena across different areas of plasma physics. We have cited theoretical works and applications to solar and magnetospheric physics. The exact onset properties of the plasmoid instability will likely be different in the different applications. However, the general pattern is seen throughout, namely a magnetic topology involving a separator which bifurcates, creating many plasmoids, on a time scale much shorter than that of  the other dynamics of the system. Even in systems that do not originally contain a separator but create one via deformation of the magnetic field \citep[e.g.][]{wyper14b}, the plasmoid instability, as we have described in this Letter, is found. In short, the magnetic topology we consider is general and its breakup is via the plasmoid instability.


Finally, we argue that the plasmoid instability will be important to consider in models of eruptive flares too. For example, an eruptive flare can be created with the same magnetic topology that we have considered with the one difference of changing the direction of the overlying coronal field \citep{dmac14}. If the active region and coronal field directions are close to anti-parallel at their boundary, stronger reconnection can occur and break the tension of the overlying field. In \cite{dmac14} an eruption occurs with both external and internal reconnection following the behaviour of the plasmoid instability, i.e. the formation of many plasmoids on a short timescale. Although the simulation of \cite{dmac14} is based on flux emergence, it falls under the class of flare models described by the breakout model \citep{antiochos99}. Manifestations of the plasmoid instability can be found in simulations of the classical breakout model. Although it is difficult to resolve plasmoids in 3D breakout simulations, they are visible in 2.5D simulations \citep[e.g.][]{macneice04,karpen12}. Indeed, \cite{karpen12} describe weak tearing followed by fast reconnection that is very similar to the reconnection pattern we have found.

As well as simulations, there is a growing body of evidence for the plasmoid instability in observations, particularly for eruptive flares in a variety of scenarios \citep[e.g.][]{takaso11, dai18, zhang19}. As the resolution of both observations and simulations increases, we conjecture that the plasmoid instability will be {manifested in many different aspects of impulsive flares.}

\section*{Acknowledgements}
LF acknowledges support by the UK Science and Technology Facilities
Council under grant number ST/P000533/1. Computational resources were provided by the EPSRC funded ARCHIE-WeSt
High Performance Computer (www.archie-west.ac.uk), EPSRC grant no. EP/K000586/1. {We both thank Dr Andrew Haynes, who originally performed the separator calculations.}

\bsp	
\label{lastpage}
\end{document}